\documentclass[12pt]{article}
\usepackage{graphicx}
\usepackage{amssymb}
\usepackage{cite}  
\newcommand{\npt}{$p_T$}

\newcommand{\Gc}{GeV/$c$}
\newcommand{\Jpsi} {J/$\psi$}

\newcommand{\nY}{$\Upsilon$}

\begin{document}

\title{Heavy Ion Physics at the LHC: What's new ? What's next ?}
\author{J. Schukraft  
\thanks{ Talk given at the 'Nobel Symposium on LHC results',
\rm Krusenberg, Sweden, 13 - 17 May 2013, to be published in
			\it Physica Scripta}}

\date{CERN, DIV. PH, 1211 Geneva 23, Switzerland}
 \maketitle

\begin{abstract}
Towards the end of 2010, some 25 years after the very first collisions of ultra-relativistic heavy ions at fixed target energies, and some 10 years after the start of operation of the Relativistic Heavy Ion Collider (RHIC), the LHC opened a new era in heavy ion physics with lead on lead collisions at $\sqrt{s_{NN}} = 2.76$ TeV. After a short reminder of the main results from lower energies, this review highlights a few selected areas where significant progress has been made during the first three years of ion operation at the LHC.

\end{abstract}


\section{Introduction}
The subject of ultra-relativistic heavy ion physics is the study of strongly interacting matter under extreme conditions of high temperature and/or high matter density. QCD predicts that at a sufficiently high energy density there will be a transition from ordinary nuclear or hadronic matter to a plasma of free ('deconfined') quarks and gluons, the 'Quark-Gluon Plasma' (QGP). The discovery and characterisation of this plasma phase is thought to require a large volume of hot/dense matter and is therefore pursued in collisions of heavy nuclei at the highest energies.

This article presents a subjective selection of results and interpretations from the first three years of the LHC heavy ion program; it is based in parts on recent reviews~\cite{Muller:2012zq, Schukraft:2011np} in which most of the relevant LHC data and references to the primary literature can be found. Theoretical aspects and implications are covered and referenced in more detail in an accompanying article~\cite{Mullerrev}.

\subsection{Prior art: main results from fixed target experiments and RHIC}

After pioneering experiments at relativistic energies (GeV/nucleon) in the 1970's in the USA (LBNL) and Russia (JINR), the quest for the Quark-Gluon Plasma took off in 1986 at the fixed target machines at CERN (SPS) and BNL (AGS), first as an exploratory program with light ions (mass $\approx 30$) and in the early 1990's with actual heavy ions (mass $\approx 200$)~\cite{Schukraft:1997mv, Antinori:2003hw, Satz:2004zd}. An appraisal of the SPS program was made in 2000~\cite{Heinz:2000bk}, based on a 'common assessment' of the results from half a dozen experiments collected and published over the preceding years. It concluded that \emph{ 'compelling evidence has been found for a new state of matter, featuring many of the characteristics expected for a Quark-Gluon Plasma'}~\cite{ CernPress}. This conclusion was based primarily on three experimental observations: the copious production of hadrons containing a strange quark ('strangeness enhancement'), the yields of low mass lepton pairs ('rho melting'), and the reduced production of $J/\Psi$ mesons ('anomalous $J/\Psi$ suppression'). 

The very high abundance of strange particles, in particular of hyperons (the omega-to-pion ratio increases by up to a factor 20 from $pp$ to PbPb!), was predicted as a consequence of QGP formation. Today it is interpreted more generally as a manifestation of statistical hadronisation from a thermalized medium, where most hadrons, not only those containing strange quarks, are created in thermal equilibrium ratios~\cite{BraunMunzinger:2003zd,Becattini:2009sc,Andronic:2009qf, Becattini:2010sk}. These ratios are governed essentially by a single scale parameter T, interpreted as a chemical freeze-out temperature. 

The invariant mass distribution of prompt low mass lepton pairs showed an enhancement in the continuum yield just below the rho/omega resonance~\cite{Specht:2007ez, Rapp:2013ema}. This was qualitatively in agreement with signals expected from chiral symmetry restoration, where the mass and/or width of hadrons  are modified in the vicinity of the QGP phase boundary (in this case the rho, observed inside the medium via its two lepton decay).

Finally, $J/\Psi$ production was found significantly suppressed in central PbPb collisions relative to expectation (eg relative to a normalisation process like Drell Yan) beyond what could be attributed to confounding cold nuclear matter  effects like changes in the nuclear parton distribution functions or hadronic final state interactions~\cite{Kluberg:2009wc,Brambilla:2010cs,Rapp:2008tf}. Dubbed 'anomalous' $J/\Psi$ suppression to distinguish it from the 'normal' one due to  cold nuclear matter, the effect was within experimental and theoretical uncertainties fully consistent with the 'smoking gun' signal predicted for deconfinement in the QGP. 

These experimental results have all stood the test of time, having been confirmed and refined subsequently at the SPS as well as at RHIC. The essence of the assessment, that there is a new state of matter  in the SPS energy range, featuring some of the hallmarks of a QGP (thermalisation, deconfinement, chiral symmetry restoration) seems however, in hindsight, today more compelling than in 2000, given for example much improved low mass lepton pair results from the SPS NA60 experiment~\cite{Specht:2010xu} and new insights in some of the processes relevant for thermal particle production and quarkonia suppression from RHIC and LHC (see below).

The initial results from RHIC were summarized and assessed in 2005, based on a comprehensive (re)analysis of the first few years of RHIC running~\cite{rhichwhitepaper}. The experiments concluded that at RHIC \emph{'a new state of hot, dense matter'} was created \emph{'out of the quarks and gluons .. but it is a state quite different and even more remarkable than had been predicted'}~\cite{BnlPress}. Unlike the  expectation, with hindsight overly naive, that the QGP would resemble an almost ideal gas of weakly coupled quarks and gluons, the hot matter was found to behave like an extremely strongly interacting, almost perfect liquid, sometimes called the sQGP (where the 's' stand for 'strongly interacting').   It is almost opaque and absorbs much of the energy of any fast parton which travels through -- a process referred to as 'jet quenching -- and it reacts to pressure gradients by flowing almost unimpeded and with very little internal friction (i.e. has very small shear viscosity) ~\cite{Jacobs:2004qv, Muller:2006ee}.

Also at RHIC, the crucial experimental results as well as the inferred characteristics of the QGP --- a 'hot, strongly interacting, nearly perfect liquid' -- stood the test of time~\cite{Muller:2012zq}. The temperature was inferred by measuring direct 'thermal' photons with an inverse slope of order 200 MeV, which leads to a (model dependent) estimate of an initial temperature of at least 300 MeV. The characterisation of the QGP as almost opaque was based on the observation of a very significant suppression, up to a factor of five, of high \npt~particles found in central, head-on nuclear collisions. This suppression of high \npt~particles, which are typically leading jet fragments, is indicative of very strong final state interactions of the scattered partons with the medium, leading to significant energy loss via elastic scattering or enhanced gluon radiation. As essential control measurements, the suppression was neither seen in d-Au reactions, therefore excluding cold nuclear matter effects as the cause, nor with colour neutral probes like direct photons, which clearly establishes the effect as due to the strong (=QCD) interaction in the final state. 

The 'ideal liquid' aspect of the QGP was based on the measurement of collective particle motions, the so called 'elliptic flow', which develops in response to initial geometrical conditions and internal pressure gradients in the nuclear overlap zone. The elliptic flow magnitude at RHIC was found to essentially exhaust the maximal possible one predicted by hydrodynamics for the given initial deformation, equivalent to the response of an ideal liquid with vanishing shear viscosity. The shear viscosity over entropy ratio, $\eta/s$ was found to be compatible with a conjectured lower bound of $ \eta/s \ge 1/4\pi $ ($\hbar = k_B = 1$), a value reached in a very strongly interacting system when the mean free path approaches the quantum limit, the Compton wavelength.

\subsection{What's new from the LHC?}
Prior to LHC,  25 years of heavy ion experimentation had already revealed a \textbf{'QGP-like'} state at the SPS and not '\textbf{the} QGP', but '\textbf{a} sQGP' at RHIC. One could consider the discovery phase for the QGP as essentially over, the qualitative characterisation as well under way, and quantitative precision measurements of QGP properties as just having started~\cite{Schukraft:2011na}. A main goal for the heavy ion program at LHC was therefore to measure with increased precision the parameters which characterise this new state of matter, making use of the particular strength of the LHC:  a powerful new generation of large acceptance state-of-the-art experiments ATLAS, CMS, ALICE, and LHCb\footnote{LHCb participates in the p-nucleus part of the heavy ion program}, and a  huge increase in beam energy with the associated larger cross sections for hard probes and higher particle density, which makes for a QGP which will be 'hotter, larger, and longer living'. And indeed, LHC made significant progress towards increasing the precision on shear viscosity (see section 4) and plasma opacity (section 5) already during the first two years of ion running. However, when dealing with QCD in the nonpertubative regime, surprises should not come as a surprise, and a number of unexpected findings at LHC have helped shed new light on some old problems or issues. Two of those will be mentioned below, relating to particle production (i.e. thermalisation, section 2) and quarkonia suppression (i.e. deconfinement, section 3). And finally the very first discovery made at LHC is discussed in section 5: the appearance of a mysterious long range 'ridge' correlation in high multiplicity $pp$ reactions. It reappeared later -- and much stronger -- in the 2012 p-nucleus run, making it of great interest, and presumably of great relevance, to hot and dense matter physics, even if it's ultimate cause and connection to similar phenomena in nuclear collisions is as of today not finally settled.

\section{Particle production}
The production of different hadronic particle species in high energy collisions is a non-pertubative process and in general parameterized in phenomenological models and event generators, requiring usually a large number of parameters. A rather more economical approach is the thermal/statistical model of hadronisation~\cite{BraunMunzinger:2003zd,Becattini:2009sc, Andronic:2009qf, Becattini:2010sk,Letessier:2005qe}, which assumes that particles are created in thermal (or phase space) equilibrium; particles of mass $m$ are essentially suppressed by a Boltzmann factor $e^{-m/T}$. In its simplest implementation, the scale parameter T, identified as the chemical freeze-out temperature, is the only free parameter of the model and defines the production of all hadronic species, together with some conservation laws (baryons, strangeness,..) and an overall normalisation constant proportional to the total particle multiplicity. For small systems (small number of final state hadrons), an additional parameter $\gamma_s$ (or, equivalently, a reduced correlation volume) has to be introduced to describe the fact that strange hadrons are suppressed compared to the grand canonical thermal expectation. The temperature parameter T is found in all high energy collisions ($pp$, $e^+e^-$, $AA$) to be about 160 MeV, while $\gamma_s$ increases from $0.5-0.7$ in $pp$ to $0.9-1$ in $AA$. The fact that most hadrons containing light and strange quarks are described by the thermal model with typically better than 10-20\% precision was considered an essential and well established fact in heavy ion collisions.

The origin of the success of the thermal model over a very large range of collision systems and beam energies is however not obvious. One of two qualitative mechanisms are usually invoked which can be summarized as i) \emph{born into (phase space) equilibrium} or ii) \emph{evolving into (thermal) equilibrium}~\cite{Stock:2007zz, Andronic:2005yp, Becattini:2010sk}. The former i) could arise for example from Fermi's golden rule, the fact that reaction rates are proportional in this case to the product of a QCD matrix element and a hadronic final state phase space factor. As many different channels (matrix elements) contribute to particle yields, phase space dominates in inclusive measurements and the most conspicuous QCD remnant is the observed suppression ($\gamma_s < 1$) of strange particles in elementary reactions ($pp$, $e^+e^-$), reflecting the higher mass of the strange quark and local strangeness conservation. This strangeness suppression is then 'somehow' lifted in large systems ($AA$), effectively replacing local strangeness conservation by average global (grand canonical) conservation. The alternative explanation ii) postulates that inelastic reactions either in the \emph{partonic phase} or in the final state \emph{hadronic phase} drive initial abundances quickly towards thermal equilibrium via detailed balance. Reaction rates, in particular those involving more than two initial state hadrons, decrease drastically with temperature and therefore inelastic reactions seize abruptly when the system expands and cools, preserving chemical equilibrium ratios and a freeze-out temperature still very close to the hadronisation transition.

Both explanations for the uncanny success of statistical models are conceivable, but difficult to underpin by quantitative dynamical calculations. And there is no fully satisfactory explanation for the strikingly similar,yet distinct, pattern of particle production in small ($pp$, $e^+e^-$) and large ($AA$) systems: In i) the mechanism of strangeness enhancement (i.e. the change in 
$\gamma_s$ or the correlation volume) remains qualitative and hand waving, whereas in ii) particle production in small systems, which would not be expected to thermalize, is usually not even addressed.
Some experimental observations are also counterintuitive: If hadronisation is sudden, from an equilibrated partonic phase (the sQGP) with little inelastic hadronic final state interactions, the particle ratios should reflect parton equilibrium, which in general would result in particle fractions very different from hadron equilibrium. If, on the contrary, particle yields are established via hadronic reactions, a common chemical freeze-out of all particle species seems a priori unlikely. One should expect to see at least \emph{some} indications for sequential freeze-out, where hadrons with large inelastic cross section stay longer in equilibrium and freeze out later (at lower temperature) than those with significantly smaller reaction cross sections. 

Measuring identified particles at LHC was nevertheless considered a somewhat boring exercise, as finding thermal particle ratios essentially identical to the ones measured at RHIC was thought to be one of the safest predictions~\cite{Abreu:2007kv}. It therefore came as quite a surprise when some particle fractions, in particular for the mundane proton, one of the most frequently produced hadrons, were found to differ considerably from expectations (and, to a lesser extent, from the ones measured at RHIC), while others, including those for multi-strange hyperons, were well in line with thermal predictions. Also the particle ratios measured in proton collisions seem less well described by thermal fits than data at lower energy, even when allowing for strangeness under-saturation ($\gamma_s < 1$)~\cite{Oeschler:2011ny}.

The increasing deviation of hadron ratios with energy in  $pp$ collisions, despite the overall increase in multiplicity which should bring results closer to statistical predictions, could fit naturally with the 'QCD plus phase space' interpretation. At the LHC, cross sections for hard processes are large and hard scattering is important even for the average minimum bias collision, so maybe semi-hard QCD processes are making their presence felt as stronger deviations from phase space dominance? 

Converging on a potential reason for the measured nuclear particle ratios at LHC proved harder, and a lively discussion is taking place centring on three explanations: i) Reduced thermal freeze-out temperature, ii) sequential hadron freeze-out, and iii) non-equilibrium parton freeze-out. Standard one parameter thermal fits (i) give a somewhat poor description of measured particle ratios and return a chemical freeze-out temperature T significantly below the one extracted previously from RHIC (by 6 to 10 MeV)~\cite{Andronic:2012dm}. Even if one accepts the LHC fits as tolerable and within the range of accuracy of the thermal model, the question why and how the chemical freeze-out temperature would come down with increasing energy remains open. Sequential models (ii) try to estimate abundance changes from inelastic hadronic final state interactions after initial chemical freeze-out~\cite{Becattini:2012xb}. In any thermally evolving system, sequential (time and temperature ordered) freeze-out of different degrees of freedom with different cross sections and mean free paths \emph{must} exist at some level; the question is one of magnitude rather than of principle. Some non-equilibrium calculations, usually done with the help of event generators or kinetic transport models, give results remarkably close to the LHC data and also improve thermal model fits at lower energies, making this mechanism a plausible explanation. However, given the multitude of unknown hadronic reaction cross sections which are needed in the final state rescattering calculations, as well as the need to consider multiprong initial states to satisfy detailed balance requirements, make these calculations more of an art than an exact science and the arguments qualitative rather than quantitative ones. Finally (iii), extensions to the standard model~\cite{Petran:2013lja}, where non-equilibrium thermodynamics (including supercooling below the phase transition) is applied to the parton rather than to the hadron phase, can well fit all currently available data, albeit at the expense of two additional free parameters.

The final resolution of the 'proton puzzle' is still outstanding, and will require a more complete set of particle ratio measurements at LHC as well as revisiting the RHIC results to confirm with better significance if particle ratios in central nuclear collisions indeed evolve with energy. Whichever explanation will finally prevail, the unexpected LHC results are a welcome fresh input likely to advance our understanding of the 'unreasonable success' of the statistical/thermal model of particle production.
  
\section{Quarkonium suppression}
Heavy flavour quarks (charm, bottom) have always been an important tool to probe the quark gluon plasma~\cite{Kluberg:2009wc,Brambilla:2010cs,Rapp:2008tf}. They are created early in the collision, by (semi)hard processes amenable to QCD calculations, and their further dynamical evolution is then modified by the surrounding medium. In particular the \Jpsi~and $\Upsilon$ families should be suppressed in heavy ion collisions in comparison with $pp$, primarily as a consequence of deconfinement ('melting') in the QGP. The magnitude of the suppression for different quarkonium states should depend on their binding energy, with strongly bound states such as the $\Upsilon$ showing less or no modification. 
 
While the 'anomalous' $J/\Psi$ suppression discovered at the SPS was considered one of the strongest indications for the QGP, the RHIC results showed essentially the same suppression at a much higher energy, contrary to most expectations and predictions from both QGP and non-QGP models. These initially very confusing results kept the interpretation of the smoking gun signal for deconfinement ambiguous for the last 10 years. A possible explanation was that the directly produced \Jpsi~is not supressed at all, neither at SPS nor at RHIC, and only the high mass charmonium states $\psi$' and $\chi_c$, which populate about 40\% of the observed \Jpsi~yield, are affected by the medium. These weakly bound high mass states should dissociate very close to or even below the critical transition temperature, but they also may be easily destroyed by hadronic final state interaction, without the need for invoking a QGP. Alternatively, it has been suggested that \Jpsi~suppression actually increases with energy but is more or less balanced by a new production mechanism, i.e. recombination at the phase boundary of two independently produced charm quarks~\cite{BraunMunzinger:2009ih}.

LHC data seems to have resolved the \Jpsi~puzzle in favour of the coalescence picture~\cite{Abelev:2012rv}. As predicted by recombination, the large charm cross section at LHC leads to \emph{less} \Jpsi~suppression at LHC, albeit not to \Jpsi~enhancement relative to pp, which was within the range of coalescence prediction and would have made the case clear cut. The suppression is also less strong at low \npt, where phase space favours recombination, in clear contrast to the opposite \npt~dependence found at SPS and RHIC. 

LHC results from the $\Upsilon$ family~\cite{Chatrchyan:2011pe} are fully consistent with the expectation from a deconfining hot medium that quarkonia survival decreases with binding energy, i.e. in terms of suppression factors: \nY(3S) $>$ \nY(2S) $>$ \nY(1S). The \nY(1S) is suppressed by about a factor of two in central collisions, the \nY(2S) by almost an order of magnitude, and only upper limits have been measured for the \nY(3S). As only about 50\% of the observed \nY(1S) are directly produced, these results are actually compatible with almost complete melting of all high mass bottonium states and survival of a lone,  strongly bound \nY(1S), which according to lattice QCD may melt only at temperatures far above the critical temperature.

While at first sight charm quark coalescence may appear as yet another process complicating and masking quarkonium deconfinement, it is actually a respectable and important deconfinement signal in itself: only in a colour conducting, deconfining 
medium can quarks roam freely over large distances ($>>$ 1 fm), and this is exactly what two charm quarks -- unlike light quarks produced independently at well separated locations and at early times -- have to do in order to combine during hadronisation.

Pending surprises from the LHC's p-Pb run, which should quantify the role of cold nuclear matter effects (shadowing/saturation and final state interactions), heavy quark diffusion and recombination is well on the way of becoming a new and accepted ingredient needed to make sense of the different patterns seen in \nY~ and \Jpsi~production over a wide range of energies. Together with, and complementary to, the original advocated process of dissociation, it may even eventually deliver on the promise of quarkonia production as an unambiguous signal of deconfined partonic matter.

\section{Elliptic flow}
The observation of robust collective flow phenomena in heavy ion reactions~\cite{Voloshin:2008dg} is arguably the most direct evidence for the creation of a strongly interacting, macroscopic (i.e. large compared to the mean free path) and dense matter system in nuclear collisions. Matter properties like the equation of state, sound velocity or shear viscosity, can be extracted by comparing measurements and hydrodynamic model calculations of elliptic (i.e. azimuth dependent) and radial (azimuthally averaged) flow. Flow depends not only on matter properties but also on initial conditions, in particular the geometrical distribution of energy density within the nuclear overlap zone and the resulting pressure gradients. In general the nuclear impact parameter, and therefore the reaction zone geometry relevant for initial conditions, can be measured rather well event-by-event for each individual collision, using global event observables like particle multiplicity or forward 'zero degree' energy. However, the remaining model dependence of the energy density profile is sufficiently large to dominate the systematic errors and limit the accuracy with which the matter parameters can be extracted from the data.  Before LHC turn-on, the defining property of the 'perfect liquid', the shear viscosity-to-entropy ratio $\eta/s$, was only known to be within a factor of about five to the conjectured quantum limit $1/4\pi$.

Azimuthally dependent collective motions are usually analysed in terms of a Fourier expansion with respect to the reaction plane, with the first order component, $v_1 \propto \cos(\varphi)$, called directed and the second order component, $v_2 \propto \cos(2\varphi)$, called elliptic flow. Higher order components were thought to be small or vanish for symmetry reasons. Between 2005 and 2010, based on observations at RHIC, suggestions were made that the geometrical overlap shape could fluctuate event-by-event, even at fixed impact parameter, because of the stochastic nature of nucleon-nucleon collisions. These fluctuations would generate 'lumpy' initial conditions which could give rise to higher harmonic Fourier components. However, these suggestions remained controversial.

When first azimuthal flow data from the LHC became available in early 2011, the evidence from all three experiments, as well as new results shown by the two RHIC collaborations, was overwhelming~\cite{Schutz:2011zz}: Fluctuations, event-by-event, of the energy density in the initial state do give rise to complex collective flow patterns, which, when analysed in terms of Fourier coefficients, are measurable and significant up to at least 6th order ($v_1, v_2, ..v_6$)! The signal strength of different harmonics, their particle mass, centrality and momentum dependence were all in excellent agreement with expectations from hydrodynamics.

The correlation patterns induced by flow fluctuations had actually been strong and clearly visible since many years also in the RHIC data; however, before 2011, they were in general not recognized as hydrodynamic in origin but discussed in terms of fancy names ('near side ridge, away side cone') and fancy explanations ('gluon Cerenkov radiation, Mach cone, ..')~\cite{Nagle:2009wr}. At LHC, the large acceptance of the experiments, together with the high particle density (as a collective effect, the flow signal increases strongly with multiplicity) made the observation and interpretation straightforward and unambiguous. 

The fact that energy density fluctuations on the scale of a fraction of the nuclear radius in the initial state are faithfully converted into measurable velocity fluctuations in the final state was a most amazing, and also most useful, discovery: One could not only identify the average, almond shaped 'face of the collision zone', but recognize much finer structures, the 'warts and wrinkles', of nuclear collisions.  The analysis of flow has been invigorated and is advancing rapidly ever since, with direct measurements of the fluctuation spectrum~\cite{Aad:2013xma}, using event-by-event measurement and selection of flow as an analysis tool~\cite{Schukraft:2012ah}, and even finding non-linear mode mixing between different harmonics~\cite{Qiu:2012uy}. Like temperature fluctuations in the cosmic microwave background radiation, which can be mapped to initial state density fluctuations in the early Universe, collective flow fluctuations strongly constrain the initial conditions (initial density distributions) and therefore allows a better measurement of fluid properties. Since 2011, the limit for the shear viscosity has come down by a factor of two ($\eta/s$ smaller than two to three times $1/4\pi$). It is now precise enough to see a hint of a temperature dependence (slightly increasing from RHIC to LHC)~\cite{Mullerrev}, and future improvements in data accuracy and hydro modelling should either further improve the limit, or give a finite value for $\eta/s$. In either case, improved precision is relevant as the shear viscosity is directly related to the in-medium cross section and therefore contains information about the degrees of freedom relevant in the sQGP via the strength and temperature dependence of their interactions.

\section{Jet quenching}
High energy partons interact with the medium and loose energy, primarily through induced gluon radiation and, to a smaller extent, elastic scattering~\cite{Majumder:2010qh}. The amount of energy lost, $\Delta E$,  is expected to depend on medium properties, in particular the opacity (density, interaction strength) and the path length L inside the medium, with different models predicting a linear (elastic $\Delta E $), quadratic (radiative $\Delta E$), and even cubic (AdS/CFT) dependence on L. In addition, $\Delta E$ also depends on the parton type via the colour charge (quark versus gluon), the parton mass via formation time and interference effects (light versus heavy quarks), and finally somewhat on the jet energy. The total jet energy is of course conserved and the energy lost by the leading parton appears mostly in the radiated gluons, leading in effect to a medium modified softer fragmentation function. 
Jet quenching (i.e. measuring the medium modified fragmentation functions) is therefore a very rich observable which probes not only the properties of the medium but also properties of the strong interaction.

Jet quenching was discovered at RHIC not with jets, which are difficult to measure in the high multiplicity heavy ion background environment, but as a suppression of high \npt~'leading' jet-fragments. While the effect was experimentally very clean and significant with suppression factors up to five, the information was also very limited, impeding a quantitative and model independent determination of matter properties or energy loss mechanisms. 

The high energy of LHC and the correspondingly large cross sections for hard processes make high energy jets easily stand out from the background even in central nuclear collisions. Jet quenching is therefore readily recognized and measured, with many unbalanced dijets or even monojets apparent in the data~\cite{Aad:2010bu}. While the amount of energy lost in the medium can be of order tens of GeV and therefore even on average corresponds to a sizeable fraction of the total jet energy, it is nevertheless close to the one expected when extrapolating RHIC results to the higher density matter at LHC. The two jets remain essentially back-to-back (little or no angular broadening relative to pp) and the radiated energy ($\Delta E$)
 is found in very low \npt~particles ($< 2$ \Gc) and at large angles to the jet direction~\cite{Chatrchyan:2011sx}. The latter two findings were initially a surprise, but are now incorporated naturally into models where the energy is lost in multiple, soft scatterings, and the radiated gluons are emitted at large angles. The parton then leaves the matter and undergoes normal vacuum fragmentation, i.e. looking like a normal $pp$ jet but with a reduced energy. 

Additional insight into the energy loss process has come from heavy flavours~\cite{ Renk:2013kva,Dainese:2013vka}.  Like at RHIC, the suppression of charm mesons is virtually identical to the one of inclusive charged particles, stubbornly refusing to show the difference expected from the stronger coupling (colour charge)  of gluons, which are the source of the majority of charged particles, compared to quarks. The mass effect however seems to be as predicted: At intermediate \npt, beauty shows less suppression than charm, whereas at very high \npt~($E/m >> 1$) b-jets and inclusive jets show similar modifications.

\section{Discoveries}
The first discovery made at LHC was announced in Sept. 2010~\cite{cmsseminar} on a subject which was as unlikely as it was unfamiliar to most in the packed audience: The CMS experiment had found a mysterious 'long range rapidity correlation' in a tiny subset of extremely high multiplicity $pp$ collisions at 7 TeV~\cite{Khachatryan:2010gv}. While in the meantime far eclipsed by the discovery of 'a Higgs-like particle', this 'near side ridge' is arguably still the most unexpected LHC discovery to date and spawned a large variety of different explanations, from mildly speculative to outright weird~\cite{Li:2012hc}. The most serious contenders are saturation physics, as formulated in the Color Glass Condensate model (CGC), and collective hydrodynamic flow. Hydrodynamics is of course a very successful framework to describe long range correlations in the macroscopic hot matter created in heavy ion reactions, but was not supposed to be applicable in small systems like $pp$ collisions, where typically only a few, or at most a few ten, particles are produced per unit of rapidity.  
The CGC is a 'first principles' classical field theory approximation to QCD which is applicable to very dense (high occupation number) parton systems like those found at small-x and small $Q^2$ in the initial state wave function of hadrons. It has been successfully used to describe some regularities seen eg in ep collisions at HERA ('geometric scaling') and to model the initial conditions in heavy ion physics.

Lacking further experimental input, no real progress was made to unravel the origin of these long range $pp$ correlations until the ridge made a robust come-back with the first LHC proton-nucleus run some two years later (p-Pb at $\sqrt{s_{NN}} = 5$ TeV~\cite{ Loizides:2013nka}). The correlation strength was actually significantly stronger than in $pp$ at the same multiplicity, and in quick succession it was discovered that: the ridge was actually double-sided, showing correlations between particles both close by in azimuth as well as back-to-back; a Fourier analysis revealed both even ($v_2$) as well as odd ($v_3$) component; the correlation strength measured  with four particles was almost identical to the one measured with two, indicating strongly a collective or at least multi-particle origin; and finally the dependence of the correlation strength on particle mass was virtually identical to the one expected from hydrodynamic flow.

All characteristics of the p-Pb ridge are very natural for and in good agreement with a hydrodynamic collective flow origin of the correlation. Even the strength of the signal and its multiplicity dependence are of the correct order of magnitude (within a factor of two) if one uses some reasonable geometrical initial conditions and a standard hydro model and just {\bf postulates} that the system, some 1 fm in size and lifetime, behaves like a macroscopic ideal fluid. Just like the matter created in central Pb-Pb collisions, which is of order 5000 $fm^3$ and therefore larger by orders of magnitude!

The question how such a tiny (few $fm^3$) system could thermalize in essentially no time, maybe even become a small serving of sQGP, has kept the case open, despite what looks like considerable evidence. However, the idea of having small sQGP fireballs created even in $pp$ collisions at the LHC may be less radical than it seems on first sight: 

At high energy, a proton is qualitatively similar to a small nucleus in some respect; an extended, finite size collection of many (sea)partons which can undergo simultaneous and independent scatterings (called multi-parton interactions in $pp$, and $N_{coll}$ in AA). Also the final state energy- or particle density (particles per unit volume) in high multiplicity $pp$ is comparable to or even larger than in central Pb-Pb. 
The thermalisation time scale commonly assumed for the sQGP created in nuclear collisions is significantly less than 1 fm/$c$~\cite{Mullerrev}; therefore the local volume required for thermalisation is presumably of comparable size, i.e. at most a few fm$^3$! Once equilibrated, the limit on $\eta/s$ implies that the mean free path in the sQGP is compatible with zero or the Compton wavelength. We do find higher harmonic flow components which originate from density fluctuations with scales of order 1 fm or less.  
There is therefore actually quite some direct and indirect evidence that in matter with initial energy densities like those produced in high multiplicity $pp$ and $pA$ (and, of course, $AA$), the 1 fm scale is big enough (and long lived enough) to approach thermal equilibrium and exhibit collective phenomena. The applicability of hydro does not depend on absolute scales (fm or km), but on dimensionless numbers, and e.g. the ratio of system size over mean free path is a big number even in a small (but extremely dense) system like high multiplicity $pp$! 

In any case, the ridge discovery in $pp$ and $pA$ at LHC is definitely more than a curiosity and likely to have profound implications for heavy ion physics, one way or another. If a sQGP (like) state can be created and studied in much smaller systems than anticipated, we can add an 'extra dimension', namely size, to our toolbox and compare $pp, pA$, and $AA$  to look for finite size effects, which may reveal information on correlation lengths and relaxation time scales not otherwise easily available\footnote[1]{we also may have to consider changing the name 'heavy ion physics'}. If, on the contrary, initial state effects and saturation physics are the answer, we would have discovered at LHC yet another new state of matter, the Colour Glass Condensate, opening a rich new field of activity for both experiment and theory. 

\section{What's next?}
As shown in selected examples above, results have come fast and on a wide range of topics for heavy ion physics during the first three years of LHC Run-1: From subtle, as yet to be fully digested hints (particle ratios) to rather suggestive and clear messages (\Jpsi~recombination), the LHC has been shining a new and very instructive light on old problems. Some properties of the sQGP have been measured with significantly better precision ($\eta/s$, opacity), improving along the way substantially our understanding of the underlying mechanism (jet quenching) or even leading to a paradigm shift (higher harmonic flow) which opened a vast new range of observables to precision experiment and precision theory. And if the strong suspicion that the surprising long range structures in $pp$ and $pA$ collisions are of collective hydro origin turns out to be correct, the 'new state of matter' will have once more shown that \emph{'  .. it is a state quite different and even more remarkable than had been predicted.'}

Some of the main experimental issues which can be addressed with nuclear beams in the upcoming runs at full LHC energy on the short to medium term are fairly clear based on the current results: 
One could hope to largely complete the measurements needed for understanding quarkonia production as a deconfinement signal and QGP thermometer. This includes quantifying 'other effects' from the past (and probably a future, high luminosity) $pA$ run, and reducing the statistical error in particular for the higher mass members of the \Jpsi~and $\Upsilon$ families as well as midrapidity low \npt~\Jpsi. Precision and sophistication in the flow analysis should further improve over the coming years, on both experimental and theory fronts, hopefully reaching a precision in e.g. $\eta/s$ of order 30\% or better, at which point the result would be precise enough for quantum corrections to the AdS/CFT lower bound to become relevant. While this is definitely a long shot, its worth aiming at because there are not many alternatives on the horizon for experimental tests of quantum string theory. And last, not least, the 'ridge puzzle' may be solved quite soon -- eg using multi-particle methods to rigorously (dis)prove collectivity -- to decide between initial state (CGC) or final state (hydro) origin. If the correlation signal turns out to be the smoking gun for saturation physics, yet another new state of matter will have been discovered at the LHC; a dense, cold, quasi-classical state of gluon matter. And even if hydro is the answer, the search should go on in $pA$ for any other sign of saturation physics. In particular in the forward direction at very low Feynman x (right in the LHCb acceptance), the conditions for saturation phenomena should be just right and proton-nucleus collisions will be the best place to look for them for some time, until an electron-ion collider comes into operation. If in fact a sensitive experimental search for the saturation physics will be carried out at the LHC, the results certainly can make, and possibly could break, the science case for such a future machine.

On a more than ten year medium- to long-term time scale, which includes experimental and machine upgrades with better detectors and higher luminosity, a comprehensive program can take place to precisely measure and test the various aspects of jet quenching, including e.g. fragmentation functions with the golden $\gamma$-jet channel and precision heavy quark measurements down to zero momentum. On this time scale one hopefully can also address signals sensitive to chiral symmetry restoration (e.g. low mass lepton pairs); a defining property of the QGP which is experimentally extremely challenging and has therefore received comparatively less attention. 

The exploration of the phases of strongly interacting matter is one of the four main pillars of contemporary nuclear physics, and one should see the LHC ion program in this broader context: The RHIC program is very active and competitive and continues to map the phase diagram at lower temperatures, in addition looking via an energy scan for the transition between normal matter and the sQGP and a 'tri-critical' point somewhere in the region at or below SPS fixed target energy. A continuation and strengthening of the SPS fixed target program is under discussion and two new low energy facilities (FAIR at GSI and NICA at JINR) are being built to study compressed matter, i.e. matter at high baryon density and (comparatively) low temperature where the phase structure may be quite different ($1^{st}$ order phase transition) and the matter is closer related to neutron stars than to the early universe.
The LHC however is and will be the energy frontier facility not only of high energy physics but also of nuclear physics for the foreseeable future, with a well-defined and extensive program and wish list of measurements. And if the first three years are any guide, strong interaction physics, while firmly rooted in the Standard Model, shows no end to surprises and discoveries and promises to keep physics with heavy ions at the LHC interesting (and fun) for quite some time to come.



\begin{thebibliography}{99}

\bibitem{Muller:2012zq}
  B.~Muller, J.~Schukraft and B.~Wyslouch,
  ``First Results from Pb+Pb collisions at the LHC,''
  Ann.\ Rev.\ Nucl.\ Part.\ Sci.\  {\bf 62} (2012) 361
  [arXiv:1202.3233 [hep-ex]].
\footnote{As far as available, reviews or overview articles -- rather than primary literature -- are cited where the reader can find further details and references.}

\bibitem{Schukraft:2011np}
  J.~Schukraft,
  ``Results from the first heavy ion run at the LHC,''
  J.\ Phys.\ Conf.\ Ser.\  {\bf 381} (2012) 012011
  [arXiv:1112.0550 [hep-ex]].


\bibitem{Mullerrev}
  B.~Muller, 
  `` Investigation of Hot QCD Matter: Theoretical Aspects,''
  To be published in the proceedings of the 'Nobel Symposium on LHC results',\rm Krusenberg, Sweden, 13 - 17 May 2013, in
{\it Physica Scripta}. 

\bibitem{Schukraft:1997mv}
  J.~Schukraft,
  AIP Conf.\ Proc.\  {\bf 531} (2000) 3.


\bibitem{Antinori:2003hw}
  F.~Antinori, A.~Billmeier and J.~Zaranek,
  ``Introduction to ultrarelativistic heavy-ion physics at the CERN-SPS,''
  AIP Conf.\ Proc.\  {\bf 631} (2003) 294.

\bibitem{Satz:2004zd}
  H.~Satz,
  ``The SPS heavy ion programme,''
  Phys.\ Rept.\  {\bf 403-404} (2004) 33
  [hep-ph/0405051].


\bibitem{Heinz:2000bk}
U.~Heinz and M.~Jacob, preprint nucl-th/0002042.

\bibitem{CernPress}
http://press.web.cern.ch/press-releases/2000/02/new-state-matter-created-cern.

\bibitem{BraunMunzinger:2003zd}
  P.~Braun-Munzinger, K.~Redlich and J.~Stachel,
  ``Particle production in heavy ion collisions,''
  In *Hwa, R.C. (ed.) et al.: Quark gluon plasma* 491-599
  [nucl-th/0304013].

\bibitem{Becattini:2009sc}
  F.~Becattini,
  ``An Introduction to the Statistical Hadronization Model,''
  arXiv:0901.3643 [hep-ph].

\bibitem{Andronic:2009qf}
  A.~Andronic, P.~Braun-Munzinger and J.~Stachel,
  ``Thermal hadron production in relativistic nuclear collisions,''
  Acta Phys.\ Polon.\ B {\bf 40} (2009) 1005
  [arXiv:0901.2909 [nucl-th]].

\bibitem{Becattini:2010sk}
  F.~Becattini, P.~Castorina, A.~Milov and H.~Satz,
  ``A Comparative analysis of statistical hadron production,''
  Eur.\ Phys.\ J.\ C {\bf 66} (2010) 377
  [arXiv:0911.3026 [hep-ph]].

\bibitem{Letessier:2005qe}
  J.~Letessier and J.~Rafelski,
  Eur.\ Phys.\ J.\ A {\bf 35} (2008) 221
  [nucl-th/0504028].
 

\bibitem{Specht:2007ez}
  H.~J.~Specht,
  ``Lepton-pair production in nuclear collisions: Past, present, future,''
  Nucl.\ Phys.\ A {\bf 805} (2008) 338
  [arXiv:0710.5433 [nucl-ex]].

\bibitem{Rapp:2013ema}
  R.~Rapp,
  ``Dilepton Production in Heavy-Ion Collisions,''
  arXiv:1306.6394 [nucl-th].

\bibitem{Specht:2010xu}
  H.~J.~Specht [NA60 Collaboration],
  AIP Conf.\ Proc.\  {\bf 1322} (2010) 1
  [arXiv:1011.0615 [nucl-ex]].


\bibitem{Kluberg:2009wc}
  L.~Kluberg and H.~Satz,
  ``Color Deconfinement and Charmonium Production in Nuclear Collisions,''
  arXiv:0901.3831 [hep-ph].


\bibitem{Brambilla:2010cs}
  N.~Brambilla, S.~Eidelman, B.~K.~Heltsley, R.~Vogt, G.~T.~Bodwin, E.~Eichten, A.~D.~Frawley and A.~B.~Meyer {\it et al.},
  ``Heavy quarkonium: progress, puzzles, and opportunities,''
  Eur.\ Phys.\ J.\ C {\bf 71} (2011) 1534
  [arXiv:1010.5827 [hep-ph]].

\bibitem{Rapp:2008tf}
  R.~Rapp, D.~Blaschke and P.~Crochet,
  ``Charmonium and bottomonium production in heavy-ion collisions,''
  Prog.\ Part.\ Nucl.\ Phys.\  {\bf 65} (2010) 209
  [arXiv:0807.2470 [hep-ph]].

\bibitem{rhichwhitepaper}
The Brahms, Phenix, Phobos, and Star Collaborations,
Nucl.\ Phys.\ A {\bf 757} (2005) 1 - 283.
 
\bibitem{BnlPress}
http://www.bnl.gov/newsroom/news.php?a=1303. 

\bibitem{Jacobs:2004qv}
  P.~Jacobs and X.~-N.~Wang,
  ``Matter in extremis: Ultrarelativistic nuclear collisions at RHIC,''
  Prog.\ Part.\ Nucl.\ Phys.\  {\bf 54} (2005) 443
  [hep-ph/0405125].


\bibitem{Muller:2006ee}
  B.~Muller and J.~L.~Nagle,
  ``Results from the relativistic heavy ion collider,''
  Ann.\ Rev.\ Nucl.\ Part.\ Sci.\  {\bf 56} (2006) 93
  [nucl-th/0602029].
 
\bibitem{Schukraft:2011na}
  J.~Schukraft,
  Phil.\ Trans.\ Roy.\ Soc.\ Lond.\ A {\bf 370} (2012) 917
  [arXiv:1109.4291 [hep-ex]].

\bibitem{Stock:2007zz}
  R.~Stock,
  ``Hadron formation in high energy elementary and nuclear collisions,''
  Int.\ J.\ Mod.\ Phys.\ E {\bf 16} (2007) 687.

\bibitem{Andronic:2005yp}
  A.~Andronic, P.~Braun-Munzinger and J.~Stachel,
  ``Hadron production in central nucleus-nucleus collisions at chemical freeze-out,''
  Nucl.\ Phys.\ A {\bf 772} (2006) 167
  [nucl-th/0511071].


\bibitem{Abreu:2007kv}
  N.~Armesto, N.~Borghini, S.~Jeon, U.~A.~Wiedemann, S.~Abreu, V.~Akkelin, J.~Alam and J.~L.~Albacete {\it et al.},
  ``Heavy Ion Collisions at the LHC - Last Call for Predictions,''
  J.\ Phys.\ G {\bf 35} (2008) 054001
  [arXiv:0711.0974 [hep-ph]].

\bibitem{Oeschler:2011ny}
  H.~Oeschler [ALICE Collaboration],
  arXiv:1102.2745 [hep-ex].

\bibitem{Andronic:2012dm}
  A.~Andronic, P.~Braun-Munzinger, K.~Redlich and J.~Stachel,
  Nucl.\ Phys.\ A904-905 {\bf 2013} (2013) 535c
  [arXiv:1210.7724 [nucl-th]].

\bibitem{Becattini:2012xb}
  F.~Becattini, M.~Bleicher, T.~Kollegger, T.~Schuster, J.~Steinheimer and R.~Stock,
  Phys.\ Rev.\ Lett.\  {\bf 111} (2013) 082302
  [arXiv:1212.2431 [nucl-th]].


\bibitem{Petran:2013lja} 
  M.~Petran, J.~Letessier, V.~Petracek and J.~Rafelski,
  arXiv:1303.2098 [hep-ph].

\bibitem{BraunMunzinger:2009ih}
  P.~Braun-Munzinger and J.~Stachel,
  arXiv:0901.2500 [nucl-th].

\bibitem{Abelev:2012rv}
  B.~Abelev {\it et al.}  [ALICE Collaboration],
  Phys.\ Rev.\ Lett.\  {\bf 109} (2012) 072301
  [arXiv:1202.1383 [hep-ex]].


\bibitem{Chatrchyan:2011pe}
  S.~Chatrchyan {\it et al.}  [CMS Collaboration],
  Phys.\ Rev.\ Lett.\  {\bf 107} (2011) 052302
  [arXiv:1105.4894 [nucl-ex]].


\bibitem{Voloshin:2008dg}
Voloshin SA, Poskanzer AM, Snellings R,
``Collective phenomena in non-central nuclear collisions,''
  arXiv:0809.2949 [nucl-ex]
\newblock {\em Landolt-Boernstein: Relativistic Heavy Ion Physics}~volume 1/23
  (Springer-Verlag, 2010), pp. 5--54, 0809.2949.

\bibitem{Schutz:2011zz}
  Y.~Schutz and U.~A.~Wiedemann,
  ``Quark matter. Proceedings, 22nd International Conference on Ultra-Relativistic Nucleus-Nucleus Collisions, Quark Matter 2011, Annecy, France, May 23-28, 2011,''
  J.\ Phys.\ G {\bf 38} (2011) 120301.

\bibitem{Nagle:2009wr}
  J.~L.~Nagle,
  ``Ridge, Bulk, and Medium Response: How to Kill Models and Learn Something in the Process,''  Nucl.\ Phys.\ A {\bf 830} (2009) 147C  [arXiv:0907.2707 [nucl-ex]]. 


\bibitem{Aad:2013xma}
  G.~Aad {\it et al.}  [ATLAS Collaboration],
  arXiv:1305.2942 [hep-ex].

\bibitem{Schukraft:2012ah}
  J.~Schukraft, A.~Timmins and S.~A.~Voloshin,
  Phys.\ Lett.\ B {\bf 719} (2013) 394
  [arXiv:1208.4563 [nucl-ex]].

\bibitem{Qiu:2012uy}
  Z.~Qiu and U.~Heinz,
  Phys.\ Lett.\ B {\bf 717} (2012) 261
  [arXiv:1208.1200 [nucl-th]].

\bibitem{Majumder:2010qh}
  A.~Majumder and M.~Van Leeuwen,
  ``The Theory and Phenomenology of Perturbative QCD Based Jet Quenching,''
  Prog.\ Part.\ Nucl.\ Phys.\ A {\bf 66} (2011) 41
  [arXiv:1002.2206 [hep-ph]].

\bibitem{Aad:2010bu}
  G.~Aad {\it et al.}  [ATLAS Collaboration],
  Phys.\ Rev.\ Lett.\  {\bf 105} (2010) 252303
  [arXiv:1011.6182 [hep-ex]].

\bibitem{Chatrchyan:2011sx}
  S.~Chatrchyan {\it et al.}  [CMS Collaboration],
  Phys.\ Rev.\ C {\bf 84} (2011) 024906
  [arXiv:1102.1957 [nucl-ex]].


\bibitem{Renk:2013kva}
  T.~Renk,
  ``Jet quenching and heavy quarks,''
  arXiv:1309.3059 [hep-ph].

\bibitem{Dainese:2013vka}
  A.~Dainese,
  ``Heavy-quark production in heavy-ion collisions,''
  J.\ Phys.\ Conf.\ Ser.\  {\bf 446} (2013) 012034.

\bibitem{cmsseminar}
http://indico.cern.ch/conferenceDisplay.py?confId=107440.

\bibitem{Khachatryan:2010gv}
  V.~Khachatryan {\it et al.}  [CMS Collaboration],
  JHEP {\bf 1009} (2010) 091
  [arXiv:1009.4122 [hep-ex]].

\bibitem{Li:2012hc}
  W.~Li,
  ``Observation of a 'Ridge' correlation structure in high multiplicity proton-proton collisions: A brief review,''
  Mod.\ Phys.\ Lett.\ A {\bf 27} (2012) 1230018
  [arXiv:1206.0148 [nucl-ex]].


\bibitem{Loizides:2013nka}
  C.~Loizides,
  ``First results from p-Pb collisions at the LHC,''
  arXiv:1308.1377 [nucl-ex].

\end{thebibliography}
\end{document}